# Room Temperature Operation of a Radiofrequency Diamond Magnetometer near the Shot Noise Limit

Chang S. Shin,[1,3] Claudia E. Avalos,[1,3] Mark C. Butler,[1,3] David R. Trease,[1,3,†] Scott J. Seltzer,[1,3] J. Peter Mustonen,[1,3] Daniel J. Kennedy,[1,3] Victor M. Acosta,[4,††] Dmitry Budker,[2,4] Alexander Pines,[1,3] and Vikram S. Bajaj[1,3,*]

(Dated: January 13, 2012)

[1]*Materials Sciences Division and* [2]*Nuclear Science Division, Lawrence Berkeley National Laboratory, Berkeley, California 94720, USA*
[3]*Department of Chemistry and California Institute for Quantitative Biosciences and* [4]*Department of Physics, University of California, Berkeley, California 94720, USA*

**Abstract:** We operate a nitrogen vacancy (NV⁻) diamond magnetometer at ambient temperatures and study the dependence of its bandwidth on experimental parameters including optical and microwave excitation powers. We introduce an analytical theory that yields an explicit formula for the response of an ensemble of NV⁻ spins to an oscillating magnetic field, such as in NMR applications. We measure a detection bandwidth of 1.6 MHz and a sensitivity of 4.6 nT/√Hz, unprecedented in a detector with this active volume and close to the photon shot noise limit of our experiment.

The negatively charged nitrogen vacancy center (NV⁻), a substitutional point defect in diamond, exhibits favorable optical and magnetic properties that have recently been exploited in several applications. For example, their brightness, optical stability, and biological inertness make NV⁻ defect-harboring nanodiamonds ideal probes in bioimaging [1] and fluorescence resonance energy transfer [2] experiments. More importantly, the NV⁻ defect forms a magneto-optical system whose spin state can be initialized and read out optically. Because the NV⁻ spin-coherence lifetimes can be as long as milliseconds in an isotopically pure diamond lattice [3], the system is an ideal platform for experimental quantum information science. Among such devices are precision magnetometers that have applications as industrial sensors, probes of magnetic materials, and as detectors of magnetic resonance. Practical magnetic field sensors for electron spin resonance (ESR), nuclear magnetic resonance (NMR), and other similar applications must sensitively detect weak, oscillating magnetic fields whose frequency and bandwidth cannot be arbitrarily controlled. Thus, transient response of an ensemble of NV⁻ centers, characterized by its sensitivity to magnetic fields oscillating over a wide bandwidth, is a critical metric for applications in NMR and magnetic resonance imaging.

To quantitatively understand the transient dynamics that limit the bandwidth in a diamond magnetometer, we have developed an analytical treatment of the transient response of an NV⁻ ensemble under microwave and optical irradiation. The NV⁻ center ground state is a spin triplet (S=1) with zero field splitting of 2.87 GHz in sublevels, i.e. $m_s = 0$ and $m_s = \pm 1$. A static magnetic field shifts the transition between $m_s = 0$ and $m_s = +1$ out of resonance with the microwave field. The system can thus be modeled as a set of three two-level systems, whose resonant frequencies are separated by ~2.1 MHz due to the hyperfine coupling between the electron spin and the $^{14}$N nuclear spin. Since the three two-level systems can be considered essentially isolated from each other [4, SI], Bloch equations can be used to model the response of each system individually, with the optically-induced spin-relaxations included in the model only through their contributions to $T_1$ and $T_2$, where $T_1$ and $T_2$ are spin-lattice relaxation time, and spin-spin relaxation time, respectively. As shown in the Supporting Information, the response time, $\tau$, of a two-level system to an oscillating field during continuous excitation is a weighted average of $1/T_1$ and $1/T_2$:

$$1/\tau = \frac{(\omega_1 \cos\phi)^2 / T_2 + \Delta^2 / T_1}{(\omega_1 \cos\phi)^2 + \Delta^2}, \qquad (1)$$

where $\Delta$ is the offset from resonance of the microwave field, $\omega_1 = \gamma B_1 / 2$ is the Rabi frequency at resonance where $\gamma$ is the electron gyromagnetic ratio, $B_1$ is the amplitude of the linearly polarized time-varying magnetic field, and the angle $\phi$ is the solution to the equation

$$\tan\phi = \frac{\Delta(1/T_2 - 1/T_1)}{(\omega_1 \cos\phi)^2 + \Delta^2}. \quad (2)$$

(In the limit where the Rabi frequency $\sqrt{\omega_1^2 + \Delta^2} \gg 1/T_1, 1/T_2$, we have $\cos\phi \approx 1$, while $\cos\phi \approx 0.5$ at low microwave power and high laser power.)

Our goal is to develop sensitive magnetometers that can be integrated with microfabricated, microfluidic NMR devices [5], in which the signal is a sum of damped oscillations with up to ~100 kHz bandwidth and ~1-3 nT amplitude, and the field from the sample dies off within 10-50 microns from the surface, limiting the detector's volume. The experiments we present below validate our analytical theory for magnetometer sensitivity and bandwidth and demonstrate that these specifications can be achieved in a practical device. We first explore the frequency response of the detector, demonstrating a bandwidth of more than 1.6 MHz. For the sensitivity experiments, we use a continuous-wave, single modulation technique in which the oscillating magnetic field itself modulates the fluorescence signal and is detected by lock-in or Fourier methods; in combination with gradiometric detection to remove technical noise, this allows us to achieve a sensitivity of ~4.6 nT/√Hz at room temperature, with an active spot size of only ~1 $\mu$m, matched to the size of small microfluidic channels and much smaller than is practical with inductive NMR detection techniques. For comparison, others have reported sensitivity of ~4 nT/√Hz using a single NV center [3] or ~20 nT/√Hz using an ensemble of NV centers, both at ambient temperatures [6]. Infrared-absorption detection, using an NV ensemble, achieved a sensitivity of ~7 nT/√Hz at 45 K [7]. In that work, a theoretical bandwidth of a few MHz was suggested, but experimental operation was limited to a few hundred Hz. With a single NV center, a detection bandwidth of a several hundred kHz has been reported [8].

The diamond sample in our experiments is S9 (NV⁻ concentration of ~2 ppm) described in Ref. [9], and the geometry of our experiment is illustrated in Fig. 1, and described further in the caption. For optically-detected ESR experiments, a static magnetic field of ~20 G was applied to break the orientational degeneracy of the NV center, and the experiments generally probed a single manifold of resolved hyperfine lines. We estimated the $T_2^*$ from the first derivative of the absorption spectra to be 130±5 ns [SI]. The frequency response of the NV ensemble at a given optical and microwave power was obtained by frequency-modulating the microwaves at a modulation rate $f_m$, centered at $f_{MW}$. The slope of the absorption spectrum was maximum about this frequency, as schematically shown in the inset of Fig. 1. Representative data together with simulations are shown in Fig. 2.

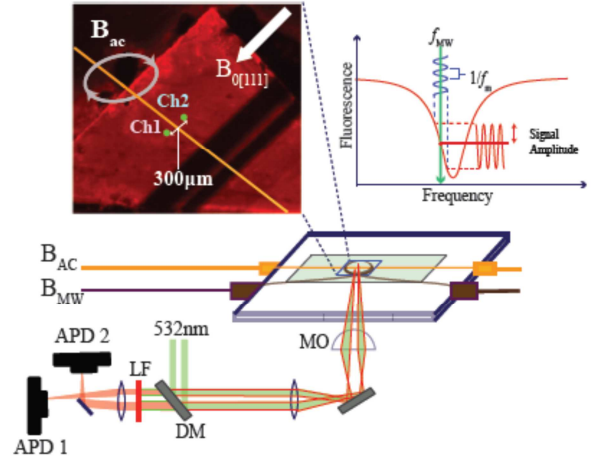

Figure 1. Schematic of experimental setup. The inset shows a fluorescence image of the diamond and a representation of the thin wire (80 $\mu$m) used to produce small oscillating magnetic fields. A three-axis set of Helmholtz coils was used to produce $B_0$, while $B_{MW}$ was generated using a high frequency microwave source. The oscillating field, $B_{AC}$ from the 80- m OD wire was produced using a function generator. MO refers to the microscope objective with NA of 0.4., LF and DM to a long-pass filter (cut-off at 650 nm) and a dichroic mirror, respectively, and APD stands for avalanche photodetector. The two excitation spots (shown in the inset) were separated by ~300 $\mu$m. The green channels refer to the 532 nm excitation, red to the fluorescence signal.

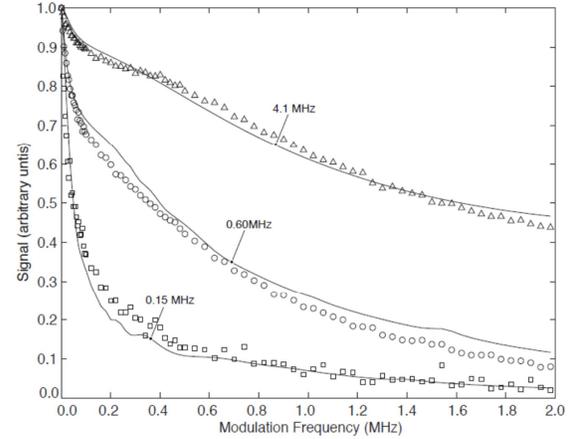

Figure 2. Representative experimental data (points) together with simulation data (thin lines) at an optical excitation power of 39 mW. Experiments were repeated at various optical powers from 0.25 to 39 mW. The signal amplitude was normalized at various Rabi frequencies from 0.06 to 4.1 MHz and plotted as a function of modulation rate.

The intrinsic $T_1$ of 462.6 $\mu$s and $T_2$ of 2.0 $\mu$s were measured by inversion recovery sequence, and Hahn-echo sequence, respectively. Further, $T_1$ and $T_2$ measurements were repeated under continuous optical excitation at various optical excitation powers, and the ensemble $T_1$ and $T_2$ were found to be decreased significantly from the intrinsic $T_1$ and $T_2$ as optical power increased, due to faster repolarization. For example, at an optical excitation power of 0.25 mW, the ensemble $T_1$ is ~152.0 µs and $T_2$ is ~1.5 µs, decreasing to 2.2 µs and 140 ns, respectively, at 40 mW of optical excitation power. While this qualitatively explains the increase in bandwidth with optical power, we note that the actual dynamics of the ensemble is complicated by the Gaussian profile of the excitation spot, which introduces a significant spatial inhomogeneity in $T_1$ and $T_2$.

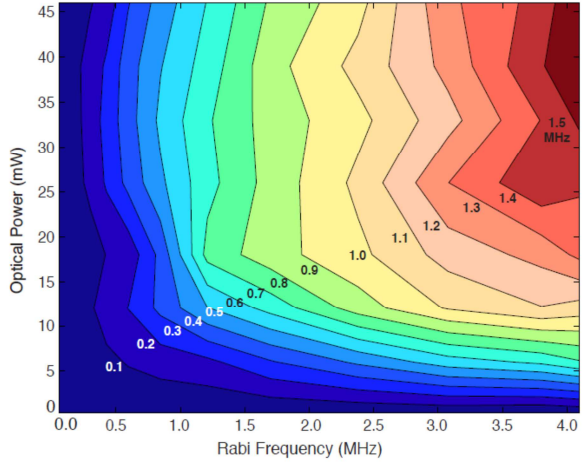

Figure 3. Experimentally estimated bandwidth is plotted as a function of optical excitation power and the microwave Rabi frequency. The system shows saturation behavior at ~12 mW of optical power, for $\omega_1$ greater than 2.0 MHz.

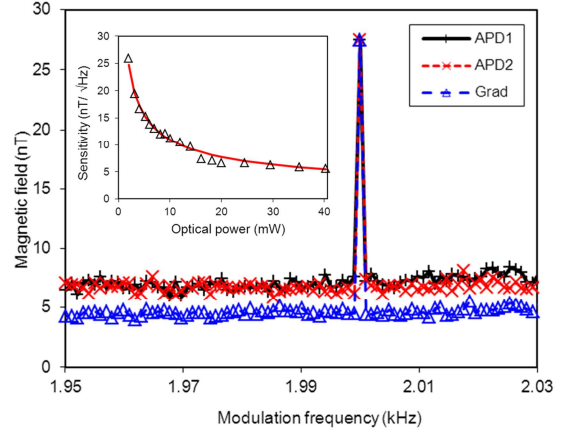

Figure 4. Magnetic field sensitivity measurements in single modulation mode, using a single fluorescence channel and a gradiometer with optical excitation power of 40.3 mW and microwave Rabi frequency of $\omega_1 = 1.6$ MHz, with one-second signal integration time. The inset shows the sensitivity as a function of optical excitation power using a single channel. Circles indicate measured data, and the solid line is a fit to Eq. (4).

Importantly, these data are explained by our analytical theory in the regime where it applies to the experiment. Bloch-equation simulations that used measured parameters are in quantitative agreement with experimental bandwidth data at low microwave powers ($\omega_1/2\pi \sim 0.1$ MHz). For laser powers near 1 mW, the simulated curves also showed qualitative agreement with the bandwidth data over the full range of microwave powers, but for laser power $\geq 6$ mW and $\omega_1/2\pi \geq 1$ MHz, simulations that used measured parameters were qualitatively different than experimental curves. The measured time constants, $T_1$ and $T_2$, changed dramatically when the laser power increased from 1 mW to 6 mW, and this suggests that the Gaussian profile of the beam introduces significant spatial inhomogeneity in $T_1$ and $T_2$ at laser powers $\geq 6$ mW. As the microwave power is increased, we can expect that the full ensemble of NV⁻ centers, inhomogeneously illuminated by the laser, will contribute to the dynamics, with each NV⁻ center having a different transient response and fluorescence intensity. As an aid to visualizing the way in which the experimental curves include contributions from NV⁻ centers that have a range of time constants, Fig. 2 shows representative experimental data, together with the simulated response of a system of two NV⁻ centers that have time constants within the expected range at the corresponding laser power of 39 mW ($T_1$=0.5 µs, $T_2$ = 0.1 µs and $T_1$ = 15 µs, $T_2$ = 1.5 µs for the individual NV⁻ spins) [SI].

In Fig. 3, we illustrate similar experiments repeated at various optical and microwave excitation powers, along with estimated bandwidths, defined as the modulation rate at which the amplitude decreases by 3 dB. At any given optical excitation power, the bandwidth increases monotonically with the microwave excitation powers, but with a larger absolute change at higher optical excitation power. For example, at an optical excitation power of 46 mW, measured before the microscope objective, the measured bandwidth increased to ~1.6 MHz as the Rabi frequency, $\omega_1$, was increased to 4.10 MHz, but at a lower optical excitation power of 0.45 mW, the bandwidth increased only to ~39 kHz when $\omega_1$ was increased over the same range.

The sensitivity of the ensemble diamond magnetometer to an oscillating magnetic field was compared for a single fluorescence channel and a gradiometer with two fluorescence channels. Microwave excitation was applied at $f_{MW}$, and a calibrated magnetic field from a thin wire was applied with a modulation rate of 2 kHz. The modulated NV fluorescence signal was integrated for 1 sec using a spectrum analyzer (Stanford Research 770) with a uniform window function.

As shown in Fig. 4, we measured a magnetic field sensitivity of ~4.6 nT/√Hz using the gradiometer and ~6.7 nT/√Hz using the single input channel. In the absence of technical noise, the sensitivity of such a magnetometer configured as a gradiometer (Fig. 1) is frequently limited by the photon shot noise [7], derived in the Supporting Information:

$$\delta B_p = \frac{4}{3} \frac{\Delta \omega}{\gamma \cdot R} \sqrt{\frac{1}{N \cdot t_m}} \qquad (3)$$

Here, $\gamma$ is the electron gyromagnetic ratio, $\Delta \omega$ is the peak-to-peak linewidth of the first derivative of the absorption spectrum, $R$ is the contrast, $N$ is the number of detected photons per unit time in the fluorescence signal, and $t_m$ is the signal integration time. Under our experimental conditions with pump power of 40.3 mW($\Delta\omega/2\pi$=7.5±0.1 MHz, $R$ ~0.043, and detected fluorescence power ~1.0 $\mu$W), this yields a photon shot-noise limited sensitivity of $\leq 4.4$ nT for 1 s integration time, approximately the same as our measured sensitivity. Further sensitivity improvements will therefore require reducing the shot-noise limit and can be expected by increasing the signal contrast using polarization-selective NV excitation [10], by improving either the detector volume or the collection efficiency [11], or by using a diamond with more favorable properties, such as longer coherence times or a higher $NV^-$:$NV^0$ ratio.

If the sensitivity is limited by the photon shot noise, then it should also depend on the optical excitation power as follows:

$$\delta B_p \propto \sqrt{\frac{1}{N}} \propto \sqrt{1+\frac{P_{sat}}{P}}, \quad (4)$$

where $P_{sat}$ is a characteristic saturation power of the NV defects, and $P$ is the applied optical excitation power [12]. The sensitivity of the $NV^-$ ensemble to a 2kHz oscillating magnetic field, was measured at various optical excitation powers from 2 to 45 mW and showed good agreement (inset in Fig. 4) with Eq. (4).

For its use in NMR, the magnetometer must detect weak, oscillating or modulated magnetic fields. To simulate this application, we employed a double-modulation technique in which we applied a frequency-modulated microwave excitation, centered at the zero crossing of the spectrum. The modulation amplitude was 4.5 MHz with modulation rates ranging from 10 to 100 kHz; we also applied an additional oscillating magnetic field of calibrated amplitude at low modulation frequencies from 10-200 Hz, designed to simulate an NMR signal [13]. In these experiments, the sensitivity improved as the first (high-frequency) modulation rate increased; this is due to the $1/f$ noise of the laser and is the principal advantage of the approach [SI], particularly where low-quality lasers are employed. We obtained a magnetic field sensitivity of 11.9 nT/√Hz using a gradiometer with this technique, limited by the additional electronic noise from the lock-in amplifier output that could not be removed with our present experimental apparatus.

In summary, we have developed an ensemble diamond magnetometer for NMR applications with a sensitivity of 4.6 nT/√Hz, exceeding by a factor ~5 the best reported sensitivity for an ensemble $NV^-$ magnetometer operating at room temperature. We measured a detection bandwidth of 1.6 MHz and developed an analytical theory to explain the transient magnetic field response of an ensemble of $NV^-$ centers under continuous microwave and optical irradiation. Our results demonstrate that magnetic sensitivity of a few nT at room temperature can be achieved with ensemble $NV^-$ centers of short $T_2^*$ ~130 ns, using modulation techniques that make the measurements robust against most of the experimental noise. The results are relevant for the development of affordable, integrated, and portable diamond magnetometers for a variety of field-sensing applications.

This work was supported by the Director, Office of Science, Office of Basic Energy Sciences, Materials Sciences and Engineering Division, of the U.S. Department of Energy under Contract No. DE-AC02-05CH11231. D.B. acknowledges support from the AFOSR/DARPA QuASAR program, IMOD, and the NATO SfP program. We thank J.F. Roch and F. Treussart for the preparation of the diamond sample and acknowledge research gifts from Agilent Technologies, Chevron Energy, and Schlumberger-Doll Research.

[†]Current address: KLA Tencor, Milpitas, CA 95035, USA
[††]Current address: Hewlett-Packard Laboratories, Palo Alto, CA 94304, USA
[*]vikbajaj@gmail.com